\documentclass[lettersize,journal]{IEEEtran}
\usepackage{amsmath,amsfonts}
\usepackage{algorithm}
\usepackage{array}
\usepackage[caption=false,font=normalsize,labelfont=sf,textfont=sf]{subfig}
\usepackage{textcomp}
\usepackage{stfloats}
\usepackage{url}
\usepackage{verbatim}
\usepackage{graphicx}
\usepackage{cite}
\usepackage{pifont}
\usepackage{multirow}
\usepackage{booktabs}
\usepackage{tcolorbox}
\usepackage{booktabs}
\usepackage{algpseudocode}
\usepackage{tcolorbox}
\usepackage{amssymb}
\usepackage{pifont}
\usepackage{makecell}
\usepackage{orcidlink}
\tcbuselibrary{skins}
\usepackage{bbding}
\usepackage{pifont}
\usepackage[table,xcdraw]{xcolor}

\newcommand{\circled}[1]{\tikz[baseline=(char.base)]{
    \node[shape=circle,draw,inner sep=0.5pt] (char) {#1};}}

\def\eg{\textit{e.g.,} }
\def\ie{\textit{i.e.,} }

\newcommand{\task}[1]{\textit{IntentRec}}
\newcommand{\ours}[1]{\textit{TreeRec}}
\newcommand{\bench}[1]{\textit{IntentRecBench}}
\definecolor{darkgreen}{RGB}{34,139,34}

\tcbset{
  my box/.style={
    enhanced,
    colframe=#1!80,
    colback=#1!10,
    attach boxed title to top left={xshift=0.2cm, yshift=-0.2cm},
    boxed title style={
      colback=#1!80,
      outer arc=0pt,
      arc=0pt,
      top=0pt,
      bottom=0pt,
    },
  },
}

\newtcolorbox{Prompt}[1]{
  my box=black,
  title=#1,
  boxrule=1.2pt,top=6pt,bottom=3.5pt,left=6pt,right=6pt
}

\begin{document}

\title{A Needle in a Haystack: Intent-driven Reusable Artifacts Recommendation with LLMs}

\author{Dongming Jin\orcidlink{0009-0002-8164-6227},~\IEEEmembership{Member,~IEEE,} Zhi Jin\orcidlink{0000-0003-1087-226X},~\IEEEmembership{Fellow,~IEEE,} Xiaohong Chen, Zheng Fang, Linyu Li, Yuanpeng He, Jia Li, Yirang Zhang, Yingtao Fang~\IEEEmembership{Member,~IEEE}


\thanks{Manuscript received April 19, 2021; revised August 16, 2021.}
\thanks{Dongming Jin, Zhi Jin, Linyu Li, Zheng Fang, Linyu Li, and Yuanpeng He are with Key Laboratory of High Confidence Software Technologies (Peking University), Ministry of Education; School of Computer Science, Peking University, Beijing, China (email: dmjin@stu.pku.edu.cn; zhijin@pku.edu.cn;)}
\thanks{Xiaohong Chen is with Shanghai Key Laboratory of Trustworthy Computing, East China Normal University, China (email: xhchen@sei.ecnu.edu.cn)}
\thanks{Jia Li and Yingtao Fang are with School of Computer Science, Wuhan University, Wuhan, China}
\thanks{Yiran Zhang is with Nanyang Technological University, Singapore}
}

\markboth{IEEE TRANSACTIONS ON SOFTWARE ENGINEERING,~Vol.~14, No.~8, August~2021}%
{Jin \MakeLowercase{\textit{et al.}}: A Needle in a Haystack: Recommending Reusable Artifacts from Intent-Level Requirements}


\maketitle

\begin{abstract}
%
In open source software development, the reuse of existing artifacts has been widely adopted to avoid redundant implementation work. Reusable artifacts are considered more efficient and reliable than developing software components from scratch. 
However, when faced with a large number of reusable artifacts, developers often struggle to find artifacts that can meet their expected needs. 
To reduce this burden, retrieval-based and learning-based techniques have been proposed to automate artifact recommendations. 
Recently, Large Language Models (LLMs) have shown the potential to understand intentions, perform semantic alignment, and recommend usable artifacts. Nevertheless, their effectiveness has not been thoroughly explored. 
To fill this gap, we construct an intent-driven artifact recommendation benchmark named \bench{}, covering three representative open source ecosystems. Using \bench{}, we conduct a comprehensive comparative study of five popular LLMs and six traditional approaches in terms of precision and efficiency. Our results show that although LLMs outperform traditional methods, they still suffer from low precision and high inference cost due to the large candidate space. Inspired by the ontology-based semantic organization in software engineering, we propose \ours{}, a feature tree-guided recommendation framework to mitigate these issues. \ours{} leverages LLM-based semantic abstraction to organize artifacts into a hierarchical semantic tree, enabling intent–function alignment and reducing reasoning time. Extensive experiments demonstrate that \ours{} consistently improves the performance of diverse LLMs across ecosystems, highlighting its generalizability and potential for practical deployment.

\end{abstract}

\begin{IEEEkeywords}
Software Reuse; Artifacts Recommendation; Large Language Models, Ontology Engineering
\end{IEEEkeywords}

\section{Introduction}
\IEEEPARstart{O}{pen-source} software ecosystems have become the backbone of modern software development. Developers rarely implement functionalities entirely from scratch. Instead, they rely on reusable artifacts such as packages and pretrained models to accelerate development and improve reliability~\cite{frakes2005software}~\cite{lim1994effects}. These reusable artifacts encapsulate mature implementations and community knowledge, enabling rapid prototyping and cost-effective maintenance. However, the flourishing open-source ecosystems provide unprecedented volumes of reusable artifacts, such as thousands of packages in the JavaScript ecosystem~\cite{wittern2016look}~\cite{web:npm} and rapidly expanding pretrained models in the HuggingFace ecosystem~\cite{jiang2023empirical}~\cite{taraghi2024deep}. This massive scale introduces a practical challenge: Given a development intent, developers must quickly and accurately identify the most suitable artifacts within a task-specific ecosystem. Therefore, \textbf{intent-driven reusable artifact recommendation (\task{})} has emerged as a critical research problem~\cite{wei2022clear}~\cite{gao2023know}, which aims to automate the mapping from high-level development intent to suitable artifacts. 

\begin{figure*}
    \centering
    \includegraphics[width=0.98\linewidth]{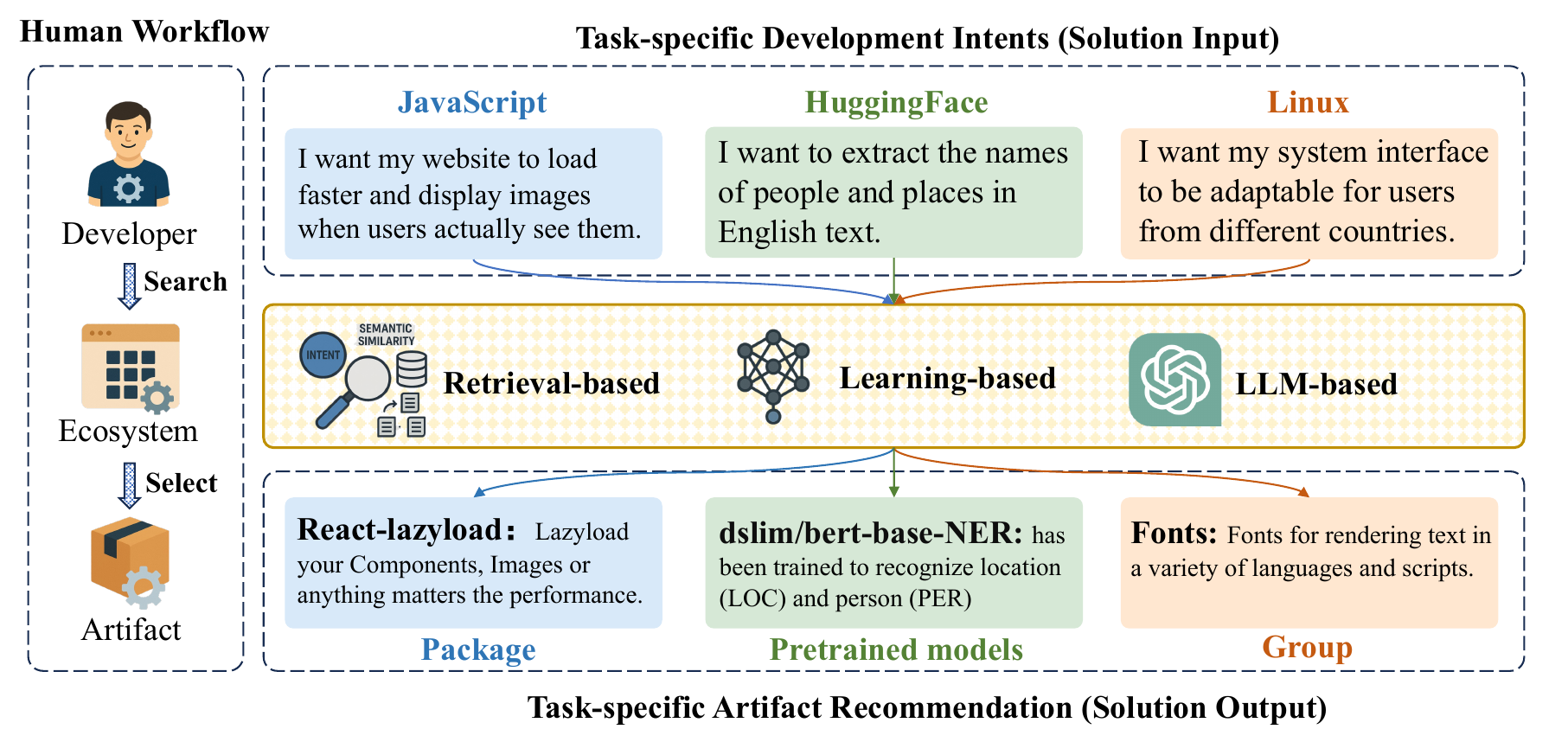}
    \caption{\textbf{The example of human workflow and alternative solution for intent-driven artifact recommendation across ecosystems.} Developers express task-specific intents, such as improving web performance, extracting named entities, or enhancing international usability. These intents can be processed through different recommendation paradigms (retrieval-based, learning-based, and LLM-based) to identify corresponding software artifacts.}
    \label{fig:teaser}
\end{figure*}

To address this challenge, various existing techniques in artificial intelligence (Figure~\ref{fig:teaser}) can be utilized. The most intuitive technique is the retrieval-based approach. They handle this task by computing textual similarities between development intent and artifact functional descriptions~\cite{gao2022propagating}~\cite{sundermann2019using}, and recommending top-$k$ artifacts with the highest similarity scores as potential candidates. However, prior work~\cite{zou2025natural} shows that matching artifacts of different abstraction levels needs more than textual similarity. 
To mitigate this issue, learning-based methods~\cite{mao2021simplex}~\cite{geng2022recommendation} can be adopted. However, they normally rely on ecosystem-specific training on huge amounts of corpora, which require unaffordable training resources (\ie labeled datasets and computational resources). 
Recently, large Language Models (LLMs) demonstrate powerful intent understanding abilities for various requirements-related tasks~\cite{jin2025iredev}~\cite{camara2023assessment} without the need for retraining or fine-tuning. However, their potential for the \task{} task has not yet been fully explored. 

\textbf{\bench{} and \textit{Multi-solution Study}.} To systematically evaluate the effectiveness of existing techniques for the \task{} task, we construct a public benchmark named \bench{}, which enables comprehensive comparison of different solutions across heterogeneous ecosystems. The \bench{} spans three types of reusable artifacts from three representative open-source ecosystems, \ie \textit{packages} in the JavaScript ecosystem\cite{frakes2005software}, \textit{pretrained models} in the HuggingFace ecosystem~\cite{ web:huggingface}, and \textit{groups\footnote{Linux projects typically bundle multiple packages that support a specific feature into a single unit called group.}} in the Linux ecosystem~\cite{jin2024first}. In total, it contains 968 samples for JavaScript, 306 samples for Huggingface, and 142 samples for Linux.  Each sample is composed of a human-written development intent description and a corresponding artifact to achieve it. To ensure intent quality and intent–artifact consistency, all samples in \bench{} are independently reviewed by a human inspection team and GPT-5, achieving 92\% and 91\% acceptance rate separately. Based on this benchmark, we conduct a comprehensive multi-solution study on six Non-LLMs approaches (\ie TF-IDF, BM25~\cite{robertson2009probabilistic}, LSI~\cite{deerwester1990indexing}, JenS~\cite{lin1991divergence}, Word2Vec, and FastText) and five popular LLMs (\eg GPT-4, DeepSeek R1, Qwen3-8B/14B/32B) from the precision and efficiency perspective. Results demonstrate that 
LLMs have significantly outperformed traditional approaches in precision still suffer from accuracy and efficiency issues. The primary limitation in accuracy arises from the abstraction gap between high-level development intent descriptions and low-level artifact functional descriptions (as shown in Figure~\ref{fig:teaser}). In terms of efficiency, computing the semantic matching score for each candidate linearly across a large number of candidates remains computationally expensive and time-consuming.

\textbf{\ours{}}. Building on the empirical findings and inspired by the feature model mechanism~\cite{hariri2013supporting}, we propose \ours{}, a hybrid recommendation framework that constructs a tree structure to capture both high-level abstraction and low-level details of each artifact and leverages this tree to guide the semantic retrieval process. It seamlessly integrates retrieval-based efficiency with the semantic reasoning capability of LLMs. During the tree construction stage, \ours{} first embeds every artifact's functional description with a pretrained model to obtain dense semantic embeddings, then clusters these embeddings to group artifacts that share common functional features, and finally prompts LLMs to summarize a high-level abstraction that becomes the parent node of each cluster. \ours{} recursively performs the embedding, clustering, and summarizing to get a hierarchical semantic tree in a bottom-up manner, which represents the artifacts at different levels of abstractions. During the recommendation stage, \ours{} traverses the constructed tree to compute a semantic relevance score between nodes and artifacts, progressively narrowing the search space until reaching the leaf nodes (\ie candidate artifacts). Finally, \ours{} prompt LLMs to re-rank the candidate artifacts to adjust their order based on their suitability for the given development intent.

We conduct extensive experiments to comprehensively evaluate the effectiveness of our \ours{}. \textbf{(1) Cross-ecosystem evaluation.} We evaluate \ours{} on all three ecosystems with GPT-4 in \bench{}. Results (Table~\ref{tab:rq1} and Table~\ref{tab:rq2}) demonstrate that \ours{} substantially improves both the efficiency and efficacy of artifacts recommendation, consistently outperforming existing baselines. \textbf{(2) Robustness evaluation.} We evaluate \ours{} with other four LLMs (DeepSeek-R1, Qwen3-8B/14B/32B) in \bench{}. Results (Table~\ref{tab:rq3}) show that \ours{} consistently boosts the performance of all evaluated LLMs across ecosystems, demonstrating strong robustness and generalizability. \textbf{(3) Ablation Study.} We conduct an ablation study of our approach by gradually adding the tree construction and re-ranking phase. Results (Table~\ref{tab:rq4}) show that both components contribute significantly. \textbf{(4) Tree Quality Evaluation.} We evaluate the quality of the constructed trees from both content and structural perspectives. Results (Table~\ref{tab:rq5}) show that \ours{} can distill artifact descriptions by approximately 85\%, generating concise yet informative high-level functional summaries and well-organized hierarchical structures. 

The contributions of this paper are summarized as follows:

\begin{itemize}
    \item We propose \bench{}, an intent-driven benchmark for reusable artifact recommendation that spans three widely used open-source ecosystems.
    \item We develop nine alternative solutions for automated artifact recommendation, spanning both traditional as well as more recent LLMs. And we conduct a a multi-solution study to empirically evaluate them on \bench{}.
    \item We propose \ours{}, a tree-guide artifact recommendation framework applicable to various LLMs to further improve their performance. 
    Qualitative and quantitative analysis show the effectiveness of our \ours{} for efficiency and efficacy improvement. 
\end{itemize}

\textbf{Data Availability}. We open-source the dataset of our \bench{} benchmark, the source code of 11 alternative solutions, and the \ours{} framework~\cite{web:code}. We hope to enable other researchers and practitioners to use it in their work on automated artifact recommendations.

\section{Background and Related Work}

\subsection{Background: Feature and Feature Model}
\begin{figure}
    \centering
    \includegraphics[width=0.99\linewidth]{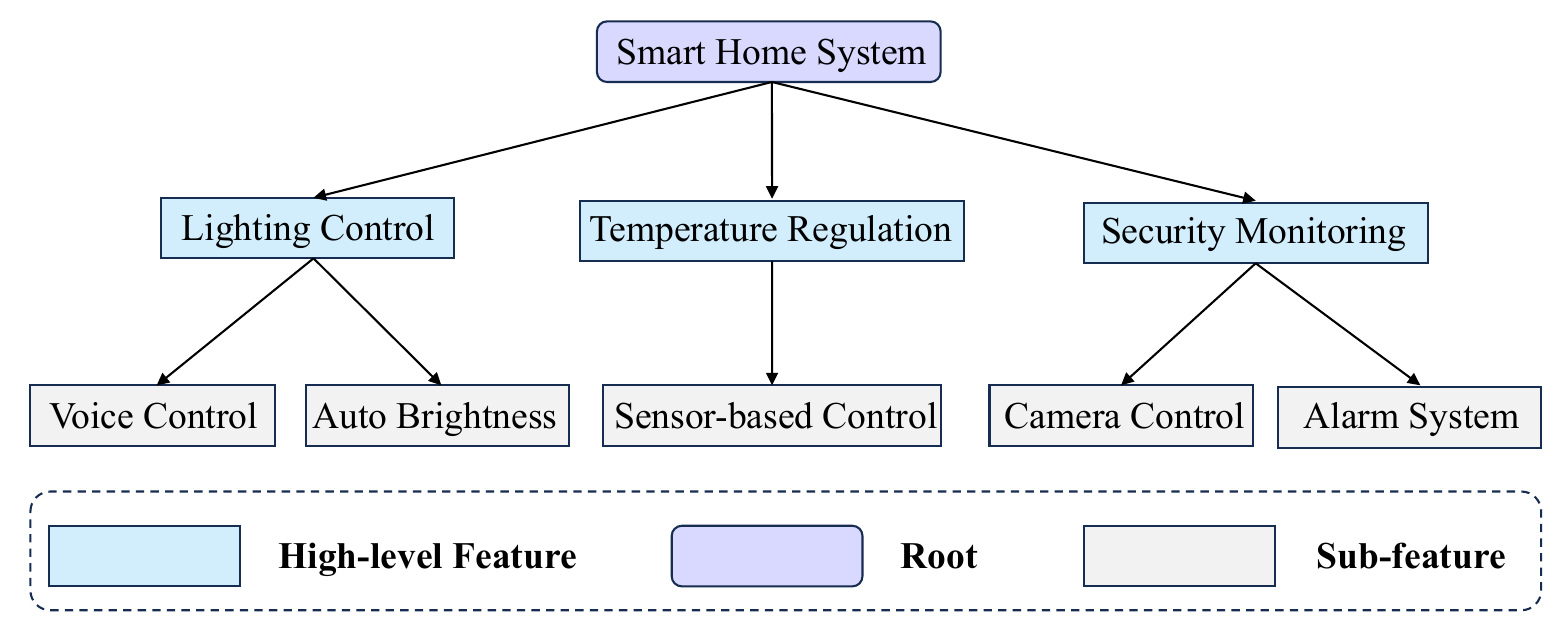}
    \caption{\textbf{A simplified example of a feature tree in a smart home system.} It illustrates how functionalities can be organized hierarchically in a feature model. High-level features such as Lighting Control, Temperature Regulation, and Security Monitoring are decomposed into representative sub-features.}
    \label{fig:background}
\end{figure}

In software engineering, especially in software product line engineering (SPLE), a feature is commonly defined as a distinctive functionality, quality, or property of a software system~\cite{kang1990feature}~\cite{pohl2005software}. It represents the core aspects of a system at the functional or quality level and can correspond to functional requirements or technical capabilities. SPLE treats features as the basic unit of variability: different products in a product family are defined by different combinations of features, and the space of feature combinations determines the scope of the entire product line. For example, as shown in Figure~\ref{fig:background}, in a smart home system, high-level features such as lighting control, temperature regulation, and security monitoring can be further decomposed into sub-features like voice-controlled switching, automatic brightness adjustment, or remote video access. Different combinations of these features form different system configurations, such as an energy-saving system or a security-prioritized system.

A feature model is a structured representation of all possible product configurations within a product line~\cite{pohl2005software}~\cite{nevsic2019principles}. It is usually expressed as a feature tree or feature diagram, which is a hierarchical semantic structure in which each node represents a feature and parent–child relationships capture ``is composed of'' semantics. Higher-level nodes describe abstract system capabilities or domain concepts, whereas lower-level nodes refine them into concrete implementation or configuration options. Continuing with the smart home example in Figure~\ref{fig:background}, the feature lighting control can be decomposed into sub-features such as manual control, voice control, and ambient-light-based automatic control. Each sub-feature can further be realized through specific modules (\eg ZigBee-based or WiFi-based lighting control). This hierarchical organization enables feature models to compactly represent both the commonality and variability of a product family, providing a unified semantic view of the system and allowing developers to clearly understand the structural differences among various configurations~\cite{pohl2005software}.

In the context of this study, the hierarchical abstraction mechanism of feature models inspires the design of our proposed \ours{} framework. Feature models organize system knowledge in a top-down manner by decomposing complex systems into multiple layers of semantic units, enabling efficient product configuration~\cite{nevsic2019principles}. Similarly, \ours{} adopts this hierarchical abstraction to organize and retrieve reusable artifacts in a bottom-up manner. 

\subsection{Reusable Artifact Recommendation.}

Reusable artifact recommendation has become a vital component in promoting practical software reuse, by helping developers discover and integrate existing software components aligned with their development needs~\cite{efthimia2025ai}~\cite{wei2022clear}. Prior research in this area can be categorized into three main directions: information retrieval (IR)-based methods, learning-based methods, and language model (LM)-based methods.

\textbf{IR-based methods.} IR-based methods primarily rely on textual similarity between user queries and artifacts descriptions. Islam et al.~\cite{islam2020socer} propose a source code recommendation technique that improves enterprise reuse by expanding natural language queries to match more relevant internal code snippets. Gao et al.~\cite{gao2023know} tackles the vocabulary mismatch problem by generating paraphrases of user questions and retrieving candidate answers from Stack Overflow posts, followed by snippet-level ranking. Silavong et al.~\cite{silavong2022senatus} introduces a fast and accurate code-to-code recommendation engine using abstract syntax trees (ASTs) and a locality-sensitive hashing variant to retrieve semantically similar code fragments. These approaches offer scalable solutions and interpretable recommendations, but often struggle with understanding deeper semantic or contextual intent behind user requirements.

\textbf{Learning-based methods.} Learning-based methods apply machine learning, particularly deep learning, to capture rich representations of code and natural language. Tao et al.~\cite{tao2022cram} models the developer’s programming context using a self-attention mechanism to suggest contextually relevant code completions. Hammad et al.~\cite{hammad2021clone} combines deep neural networks with clone detection to recommend idiomatic code fragments that frequently recur across projects. Wen et al.~\cite{wen2022code} uses a joint embedded attention network to align natural language queries with code tokens in a shared semantic space, enabling more accurate retrieval. Abid et al.~\cite{abid2021facer} leverages project-level context by monitoring feature-level progress and recommending missing features implemented in similar historical projects. Nguyen et al.~\cite{nguyen2021persona} brings in personalization, learning from individual developer history to recommend code consistent with their habits, such as naming conventions or preferred libraries. These approaches improve semantic alignment and adaptability of recommendations but generally require significant labeled data and training resources.

\textbf{LM-based methods.} LM-based approaches leveraging pretrained transformer models to bridge the gap between high-level user intent and reusable software artifacts. Sun et al.~\cite{sun2023code} embeds both queries and code snippets into the same vector space for semantic similarity-based recommendation, without the need for domain-specific supervision. Majidadeh et al.~\cite{majidzadeh2024multi} fine-tune pretrained models on traceability and matching tasks, enabling effective mapping between requirements and code even with limited training data. Recently, Prompt-based applications of generative LLMs such as GPT-3.5 or GPT-4 are also being explored~\cite{hassine2024llm}, showing potential in interpreting user requirements and recommending artifacts in a zero-shot or few-shot setting.

\subsection{Feature Tree Construction}
The feature tree can represent the commonality and variability among reusable software artifacts~\cite{hariri2013supporting}. It can organize the artifacts in a hierarchical structure based on their functionality~\cite{reiser2007multi}. There have been various studies on extracting requirements features~\cite{guzman2014users}~\cite{ferrari2013mining}~\cite{kumaki2012supporting}~\cite{acher2012extracting}. Guzman et al.~\cite{guzman2014users} used part-of-speech tagging to extract functional features from app store reviews, helping developers analyze user feedback and identify high-frequency features. Ferrari et al.~\cite{ferrari2013mining} proposed a method based on natural language processing and comparative analysis to automatically extract commonalities and variabilities from documents of competing products. Kumaki et al.~\cite{kumaki2012supporting} proposed a technique based on the vector space model to automatically analyze the commonalities and variabilities of the requirements and structural models of legacy software assets. Mathieu et al.~\cite{acher2012extracting} used the domain-specific language VarCell to extract feature models from tabular requirements descriptions, ensuring that the generated models accurately reflect the commonalities and variabilities between artifacts. However, these existing works~\cite{ferrari2013mining}~\cite{weston2009framework} primarily focus on extracting features from requirement documents for forward reuse. They have limited attention to reverse extraction from reusable artifact descriptions. In addition, they typically rely on traditional natural language processing techniques, \eg part-of-speech tagging and term weighting. These methods may lack sufficient performance. Thus, this work aims to explore more advanced techniques to extract features from reusable artifacts for reverse engineering.

\section{\bench{} benchmark} ~\label{sec:benchmark}
In this section, we first show an overview of \bench{} and then describe its construction pipeline and statistics. 

\subsection{Overview}
\bench{} aims to facilitate the development and evaluation of automated intent-driven artifact recommendation techniques. To ensure the diversity and generality of the benchmark, \bench{} spans three representative open-source ecosystems, \ie packages in JavaScript, pretrained models in HuggingFace, and groups in Linux. Each sample in \bench{} consists of two main components. \ding{182} \textbf{Intent:} A development intent description detailing the desired software feature. \ding{183} \textbf{Artifact:} A corresponding reusable software artifact that fulfills the development intent. 

\subsection{Benchmark Construction Pipeline}
\begin{figure}
    \centering
    \includegraphics[width=0.99\linewidth]{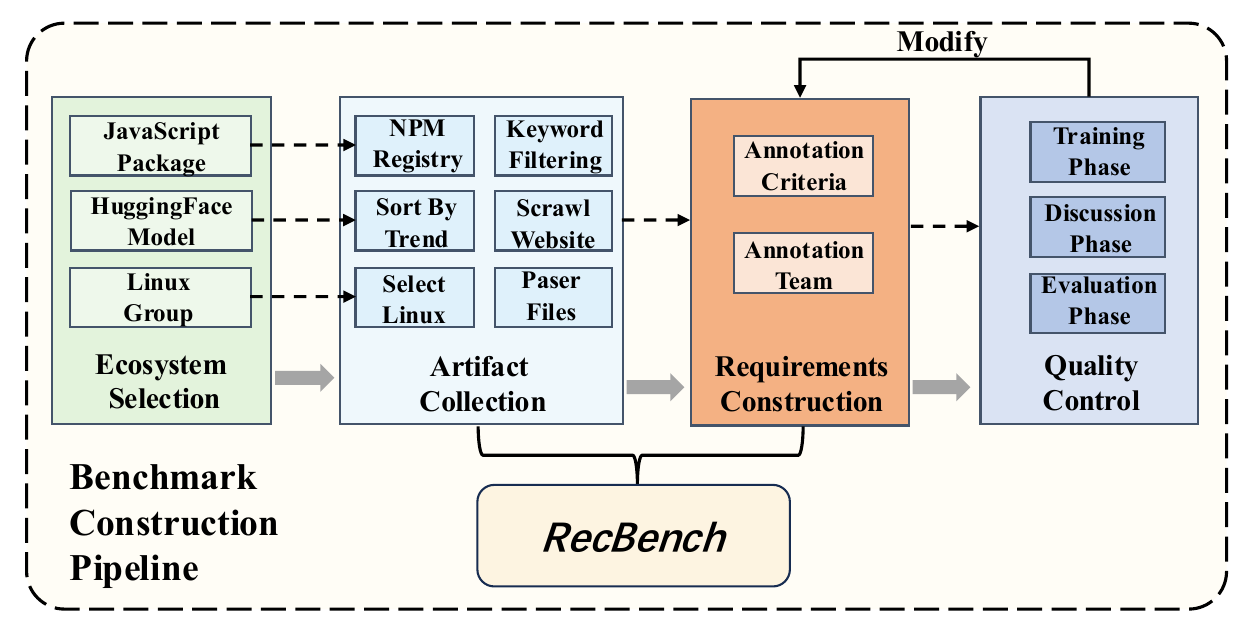}
    \caption{Overview of \bench{} Construction Pipeline}
    \label{fig:benchmark_construction}
\end{figure}

Figure~\ref{fig:benchmark_construction} illustrates the procedure of constructing \bench{}. We follow four steps to create \bench{}: (i) select open-source ecosystems; (ii) collect artifacts from ecosystems; (iii) construct intent-artifact pairs; (iv) conduct multi-stage quality control and evaluation.

\textbf{Open-Source Ecosystem Selection.} To ensure broad applicability and representativeness, \bench{} covers three prominent open-source ecosystems spanning diverse artifact types and application domains. The JavaScript ecosystem is one of the most widely adopted software package ecosystems in modern web system development, containing millions of \textit{package} artifacts across various web application scenarios (\ie Front-end, back-end, CSS, and IoT). The HuggingFace ecosystem is a major artificial intelligence community sharing \textit{pretrained model} artifacts for various NLP and vision applications. The Linux distributions ecosystem provides \textit{group} artifacts, which bundle functional modules in system-level environments to deliver specific features. Therefore, these ecosystems differ significantly in artifact types, artifact granularity, and usage contexts, which enables a comprehensive evaluation of recommendation techniques across various scenarios. Moreover, all three ecosystems provide publicly accessible and well-documented data sources conducive to collection and annotation.

\textbf{Artifacts Collection.} The step aims to construct the candidate pools of artifacts for each ecosystem, which simulates the search environment humans encountered and serves as the foundation for intent-artifact pair construction. To achieve this goal, this step involves crawling and extracting artifact metadata (\ie the name and functional description) from these selected ecosystems. Each ecosystem requires a tailored collection strategy due to its distinct structure, data accessibility, and artifact organization. We summarize the collection details for JavaScript, HuggingFace, and Linux as follows. For the JavaScript ecosystem, we utilize the official npm\footnote{NPM is a large public database of JavaScript software packages and the meta-information surrounding it.} registry API to collect the metadata of packages. Specifically, we first select twelve popular domain-relevant keywords from the official website, including front-end, back-end, CLI, documentation, CSS, testing, IoT, coverage, mobile, frameworks, robotics, and math. Then we download the metadata of packages in each keyword and fetch their monthly download count using the npm registry API. Last, to ensure data quality and reduce noise from rarely used packages, we sort the packages based on their download count and select the top 1,000 packages for each keyword. In total, we collect 9,729 packages for the JavaScript ecosystem due to the fact that several selected keywords contain fewer than 1,000 available packages. For the Huggingface ecosystem, we first sort all pretrained models on the official HuggingFace Model Hub by trending status and crawl the names of all models in the top 50 pages\footnote{The official website currently only exposes the top 50 pages.}, which obtains a list of 3,000 model names. Then, we visit the detailed page of each model in this list and download its functional description. Last, we obtain 3,000 unique models with their corresponding names and functional descriptions. For the Linux ecosystem, we first select five representative RPM-based Linux distributions following prior work~\cite{jin2024first}, \ie Fedora, CentOS, OpenEuler, Anolis, and OpenCloudOS. Then we download their source code from official or widely available mirrors, locate figuration files containing the metadata of groups and scrape these files to extract the name and functional description of each group. Finally, we merge and duplicate all groups from all five Linux distributions, which results in a list of 142 unique groups.

\textbf{Intent-Artifact Pair Construction.} This step aims to create high-quality intent–artifact pairs from the collected candidate pools to serve as a benchmark for evaluating intent-driven recommendation techniques. Given that it is unnecessary to construct intent–artifact pairs for every collected artifact, we adopt a \textit{sample-then-annotation} strategy. \textbf{(1) Sampling.} We sample approximately 20\% of artifacts from the candidate pools. Specifically, we employ a domain- and popularity-based sampling strategy for the JavaScript and HuggingFace ecosystems, while including all group artifacts from the Linux ecosystem due to its relatively small size. For instance, the JavaScript ecosystem covers domains such as front-end, back-end, testing, and IoT, whereas the HuggingFace ecosystem spans NLP, vision, and multimodal models. Within each domain, we further consider artifact popularity measured by download counts. The rationale behind this sampling strategy is to ensure that the constructed benchmark captures the functional diversity of software ecosystems and the practical relevance of artifacts that developers frequently encounter in real-world development scenarios \textbf{(2) Annotation.} In this phase, annotators compose development intent descriptions for the sampled artifacts to construct the intent–artifact pairs. Each annotation aims to describe what a developer would search for when looking for an artifact with similar functionality. To ensure annotation quality and consistency, we recruited two Ph.D. candidates and three undergraduate students in computer science. All of them are fluent English speakers and have practice with at least one of our selected three ecosystems. Before the annotation began, all annotators received detailed task instructions and reference examples to align their understanding of annotation objectives and quality criteria. During annotation, the metadata of each artifact (\ie name and functional description) was presented to the annotators. Annotators were instructed to compose corresponding intent descriptions from their perspective by following the guidelines below:

\begin{itemize}
    \item \textbf{Perspective.} Write from the viewpoint of a typical developer, focusing on the intended functionality and expected usage scenario of the artifact. 
    \item \textbf{Conciseness.} Keep the intent clear, concise, and unambiguous, avoiding technical implementation details and redundant expressions. 
    \item \textbf{Relevance.} Ensure strong correspondence between the intent and the artifact’s metadata (name and functional description). 
    \item \textbf{Quality.} Review each intent for linguistic consistency and completeness before submission. 
\end{itemize}

In total, the annotation team spent over 750 person-hours constructing 1416 high-quality intent–artifact pairs across the three ecosystems, including 968 for JavaScript, 306 for HuggingFace, and 142 for Linux.

\textbf{Quality Control.} To ensure the reliability of \bench{}, we conducted a three-phase quality assurance process, including a training phase before annotation, an expert review phase during annotation, and a quality evaluation phase after annotation. In the training phase, all annotators were assigned the same 30 artifacts to write intent descriptions. The first author evaluated the semantic equivalence among their annotations, calculated the equivalence rate, and facilitated a disagreement discussion to refine the annotation criteria. 
This process was repeated until the annotators produced semantically consistent results. During the annotation phase, the three undergraduate students performed the annotation individually. Given the time and resource constraints, not all samples could be cross-reviewed. Thus, the two Ph.D. students supervised the process and organized discussions whenever annotators encountered ambiguities or domain-specific issues. In the final quality evaluation phase, we randomly sampled 10\% of the annotated pairs (\ie 140 samples) for human expert review. 
The first author examined their semantic correctness, completeness, and relevance, achieving an acceptance rate of 92\% (\ie 129/140), which underscores the high reliability and usability of \bench{}. To complement human evaluation and further validate annotation quality at scale, we additionally employed LLMs (\ie GPT-5) to perform automated consistency and relevance assessments on the entire dataset (\ie 1416 samples). 
Specifically, GPT-5 was prompted to evaluate whether each intent accurately describes the corresponding artifact’s functionality and whether the intent–artifact pair maintains semantic alignment. Table~\ref{tab:eval_data} shows the evaluation results with GPT-5. Results show \bench{} achieves a high acceptance rate (\ie 91\%) across all ecosystems, indicating that the vast majority of pairs are semantically coherent and relevant. This guarantees fair and consistent evaluation for future studies using our \bench{}. Furthermore, the LLM’s assessment results were compared with expert review outcomes on the manually evaluated samples, achieving a consistency rate of 0.87, which demonstrates the reliability of the LLM-based evaluation.

\subsection{Benchmark Statistics}
Table~\ref{tab:bench_statis} provides a summary of the statistics for \bench{}. It reports the sample counts and the text-length characteristics of the two components (\ie \textit{Intent Description} and \textit{Artifact Functional Description (AFD)}) across the three ecosystems. For each component, we report the average, maximum, and minimum number of English words. The statistics reveals substantial variability in requirements complexity, artifact granularity, and overall data scale. 

\begin{table}[]
    \centering
    \caption{The Quality of the Constructed Dataset.}
    \begin{tabular}{ccccc}
\toprule
                 & \textbf{JavaScript} & \textbf{Huggingface} & \textbf{Linux} & \textbf{Total} \\ \midrule
Human Evaluation &  97\%&             88\%&       90\%&       92\%\\
GPT-5 Evaluation &            95\%&             87\%&       91\%&       91\%\\ \midrule
Consistency      &            0.95&             0.81&       0.86&       0.87\\ \bottomrule
\end{tabular}
    \label{tab:eval_data}
\end{table}

\begin{table}[]
    \centering
    \setlength{\tabcolsep}{5pt}
    \caption{The statistics of the \bench{} benchmark}
    \begin{tabular}{llllllll}
\toprule
\multirow{2}{*}{\textbf{Ecosystem}} & \multirow{2}{*}{\textbf{\#Samples}} & \multicolumn{3}{c}{\textbf{Length of Intent}}   & \multicolumn{3}{c}{\textbf{Length of AFD}}   \\
                                    &                                     & \textbf{Aver} & \textbf{Max} & \textbf{Min} & \textbf{Aver} & \textbf{Max} & \textbf{Min} \\ \midrule
JavaScript                          &968& 90&167& 49&17&54&2\\
HuggingFace                         &    306&  131&     48&  331& 28& 68& 11\\
Linux                               &      142                               & 88& 124& 44& 20& 43&3\\ \midrule
Overall                             &1416& 99& 331&44&20&68& 2             \\ \bottomrule
\end{tabular}
    \label{tab:bench_statis}
\end{table}
\section{Comparative Study}~\label{sec:comparative}
This section begins by defining the formulation of automated intent-driven artifact recommendation, followed by a comparative study of three types of alternative solutions. 

\subsection{Problem Formulation}
Let $L=\{a_{1},\dots ,a_{n}\}$ be an candidate artifact pool, where each artifact $a_i$ is characterised by a name $n(a_i)$ and a natural language functional description $d(a_i)$. Given a development intent description $R$ and a user-specific cut-off $k$, the artifact recommendation task is to learn a mapping
\[
G:\,(L,R,k)\;\longrightarrow\;\langle a_1,\dots ,a_k\rangle, a_j\in L,
\]
The mapping must satisfy, for any $j<j'$,
\[
\operatorname{suit}\,\!\bigl(R,\,d(a_j\bigr)\;
\ge\;
\operatorname{suit}\,\!\bigl(R,\,d(a_{j'})\bigr),
\]
where $\operatorname{suit}(\cdot,\cdot)$ measures the usage suitability
between a requirement and an artifact description. Therefore, 
the resulting list $\operatorname{Top\text{-}k}(R)=\langle a_1,\dots ,a_k\rangle$ contains the $k$ artifacts most suitable for fulfilling the requirement $R$.

\subsection{Comparative Solutions}
To comprehensively evaluate the performance of different paradigms for automated artifact recommendation, we design and compare seven alternative solutions. These solutions fall into two major categories: Non-LLM solutions and LLM-based solutions. Each solution reflects a distinct mechanism for capturing the semantic relationship between development intents and artifact descriptions.

\textbf{(i) Non-LLM Solutions.} The non-LLM category includes six alternatives that compute intent–artifact similarity through handcrafted, statistical, or shallow learned representations. \textit{Solution \circled{1} (TF-IDF).} adopts the classical term frequency–inverse document frequency representation to encode both intents and artifacts as sparse weighted vectors and ranks artifacts by cosine similarity. \textit{Solution \circled{2} (BM25)} improves over TF-IDF by introducing document length normalization and term-frequency saturation, producing more robust matching for long or unbalanced artifact descriptions. \textit{Solution \circled{3} (LSI)} performs latent semantic indexing based on singular value decomposition to project the intent and artifact representations into a shared low-dimensional latent space, thereby uncovering hidden semantic correlations. \textit{Solution \circled{4} (JenS)} represents intents and artifacts as probability distributions derived from normalized TF-IDF vectors and measures their similarity using the Jensen–Shannon divergence. This approach captures distributional differences but remains limited to shallow semantics. \textit{Solution \circled{5} (Word2Vec)} and \textit{Solution \circled{6} (FastText)} generate dense embeddings from large text corpora to capture contextual and morphological similarity. Intent and artifact representations are obtained by averaging their word embeddings and compared using cosine similarity.

\textbf{(iii) LLM-based solution.}
Solution \circled{7} leverages recent advances in large language models (LLMs) to directly infer the semantic suitability between development intents and candidate artifacts without additional training. We evaluate five representative LLMs of diverse architectures and scales (\ie GPT-4, DeepSeek-R1, Qwen3.1-8B, Qwen3-14B, and Qwen3 32B) under a unified zero-shot setting. Considering the large number of artifacts in the candidate pool, we adopt a two-stage (\ie scoring-then-recommendation) inference strategy to ensure scalability and maintain reasoning quality. In the scoring stage, each artifact is individually paired with the intent and evaluated by the LLM, which assigns a semantic relevance score ranging from 0 to 100. The higher score indicates more relevance between a development intent and an artifact. This stage finally produces a numeric relevance distribution over all artifacts.
In the recommendation stage, we select the top-scored subset (typically the top 10\%) as candidates and prompt the LLM again to generate the final top-$k$ recommendations through comparative reasoning over the selected subset.
This staged design significantly reduces the input context length while preserving semantic precision across a large artifact space.
The detailed scoring and recommendation prompts for Solution~\circled{7} are provided in our open-source package~\cite{web:code}.

\subsection{Empirical Results}

\begin{figure*}
    \centering
    \includegraphics[width=0.98\linewidth]{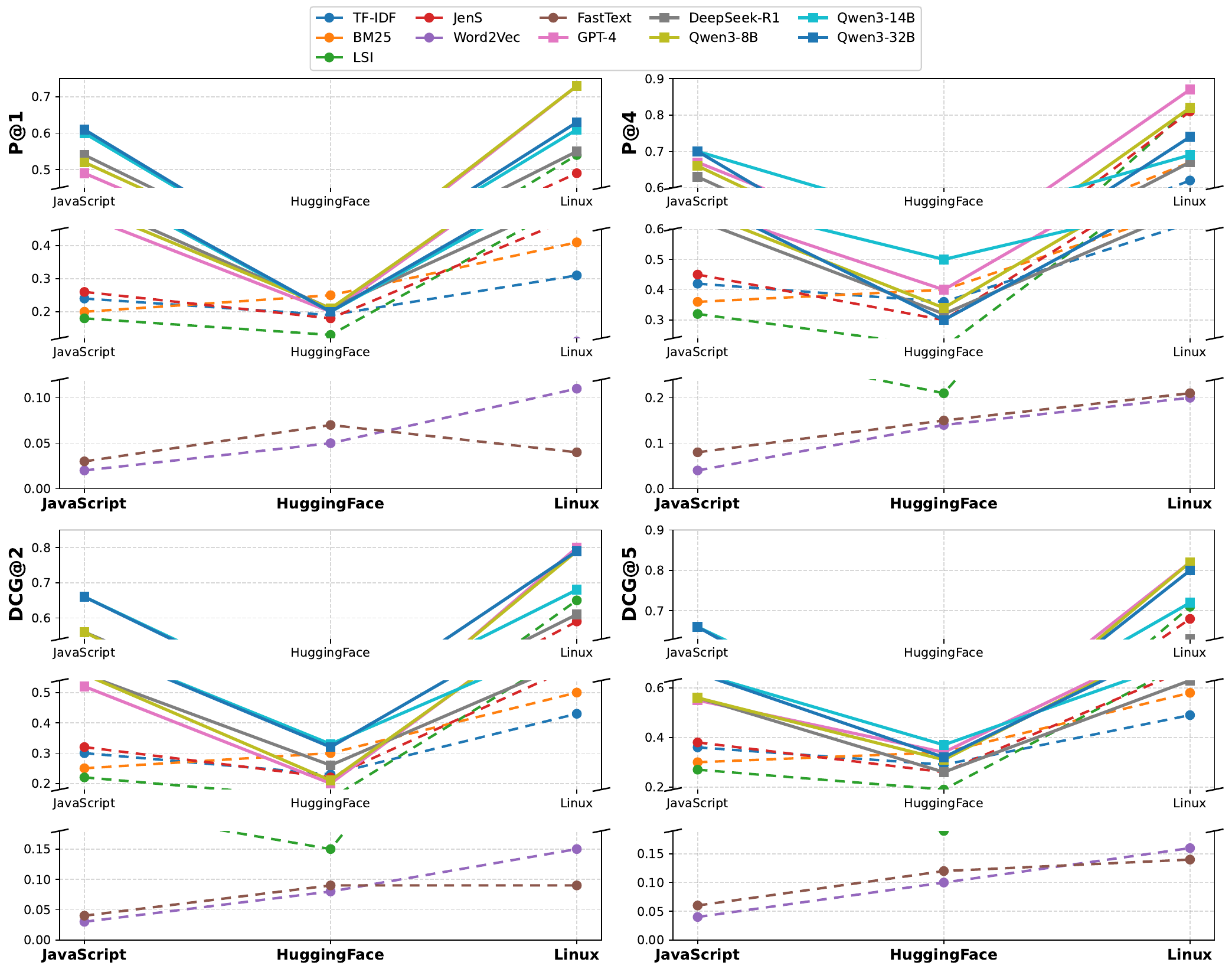}
    \caption{The empirical results for all solutions from the precision perspective.}
    \label{fig:e1}
\end{figure*}

\textbf{Precision Perspective.} Figure~\ref{fig:e1} demonstrates the precision-oriented results of the non-LLM solutions and LLM-based solutions in terms of P@1, P@4, DCG@2, and DCG@5 (Section~\ref{sec:metrics}).
\textbf{(1) Non-LLM methods exhibit stable but limited performance.} Across JavaScript and HuggingFace, TF-IDF, BM25, LSI, and JenS deliver moderate but consistent scores, reflecting their reliance on lexical or shallow semantic similarity. Word2Vec and FastText remain weak because simple embedding averaging cannot capture the richer contextual cues present in intent descriptions.
\textbf{(2) LLM-based solutions dramatically outperform non-LLM baselines.} GPT-4 achieves the strongest overall performance (e.g., P@1 of 0.49 on JavaScript and 0.73 on Linux), far exceeding all non-LLM methods. Qwen3-14B and Qwen3-32B also surpass traditional baselines on HuggingFace and Linux, showing the benefit of deeper semantic reasoning.
\textbf{(3) LLM-based solutions do not always dominate.} Despite their scale, Qwen3-14B and Qwen3-32B perform poorly on JavaScript (P@1=0.07), suggesting that model size alone cannot overcome noisy, heterogeneous artifact descriptions. The semantic gap created by such ecosystems limits their effectiveness regardless of parameter count.
\textbf{(4) Linux ecosystem is significantly easier for all methods.} All approaches perform substantially better in Linux, where concise and standardized package descriptions reduce ambiguity and allow even simple similarity-based models to match intents effectively.

\begin{figure}
    \centering
    \includegraphics[width=0.9\linewidth]{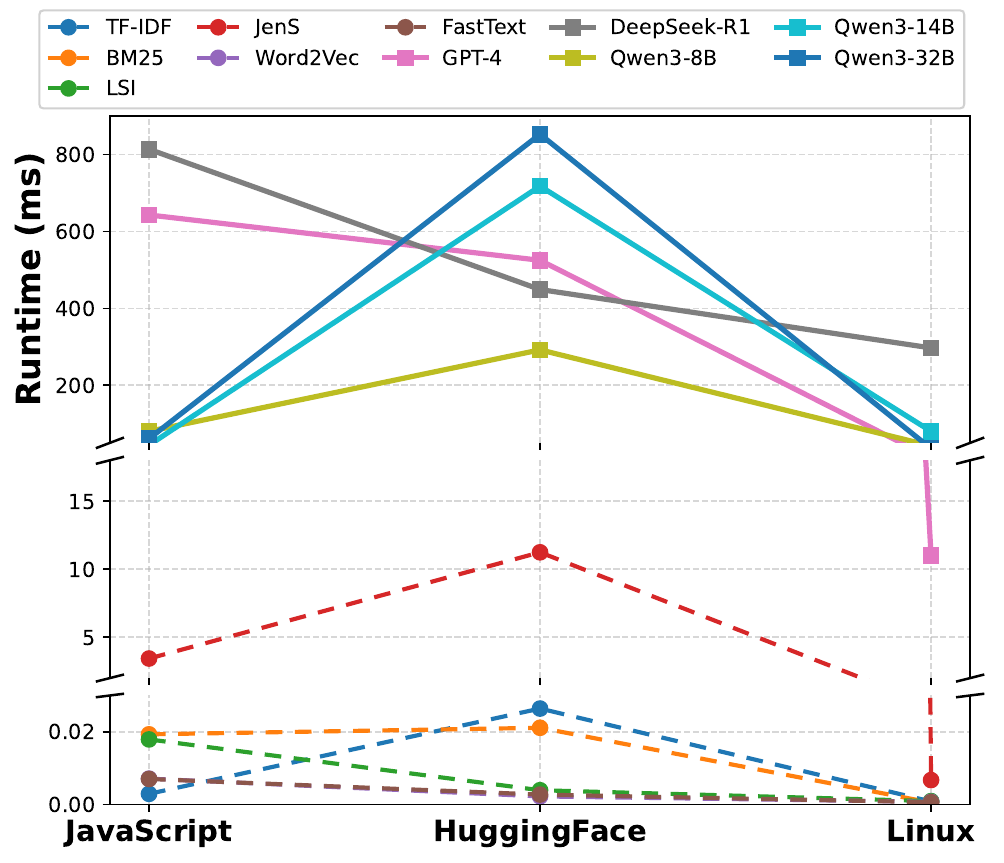}
    \caption{The empirical results for all solutions from the efficiency perspective.}
    \label{fig:e2}
\end{figure}

\textbf{Efficiency Perspective.} Figure~\ref{fig:e2} demonstrates the efficiency results of all solutions in terms of the average time cost per query.
\textbf{(1) Non-LLM solutions are extremely fast.} TF-IDF, BM25, LSI, Word2Vec, and FastText complete queries within 1–20 milliseconds, offering near-instantaneous performance suitable for large-scale retrieval; JenS is slower only due to divergence-based computations.
\textbf{(2) LLM solutions incur substantial time overhead.} All tested LLMs require tens to hundreds of seconds per recommendation, with GPT-4 and Qwen models sometimes exceeding 600 seconds. This cost arises from the need to score each artifact individually, causing runtime to scale linearly with the artifact pool.
\textbf{(3) Precision–efficiency trade-off is structural.} Non-LLM methods provide excellent speed but limited precision, whereas LLMs offer stronger semantic alignment at the expense of prohibitive latency. This gap reveals the need for a scalable design that retains LLM-level quality without linear-time matching.

\section{Approach}
This section introduces our tree-based artifact recommendation framework, named \ours{}. We begin with an overview of the framework, and then describe the details of its three constituent phases, including tree indexing, tree‑guided search, and refined re-ranking. 



\subsection{\ours{} Overview}

\begin{figure*}
    \centering
    \includegraphics[width=0.98\linewidth]{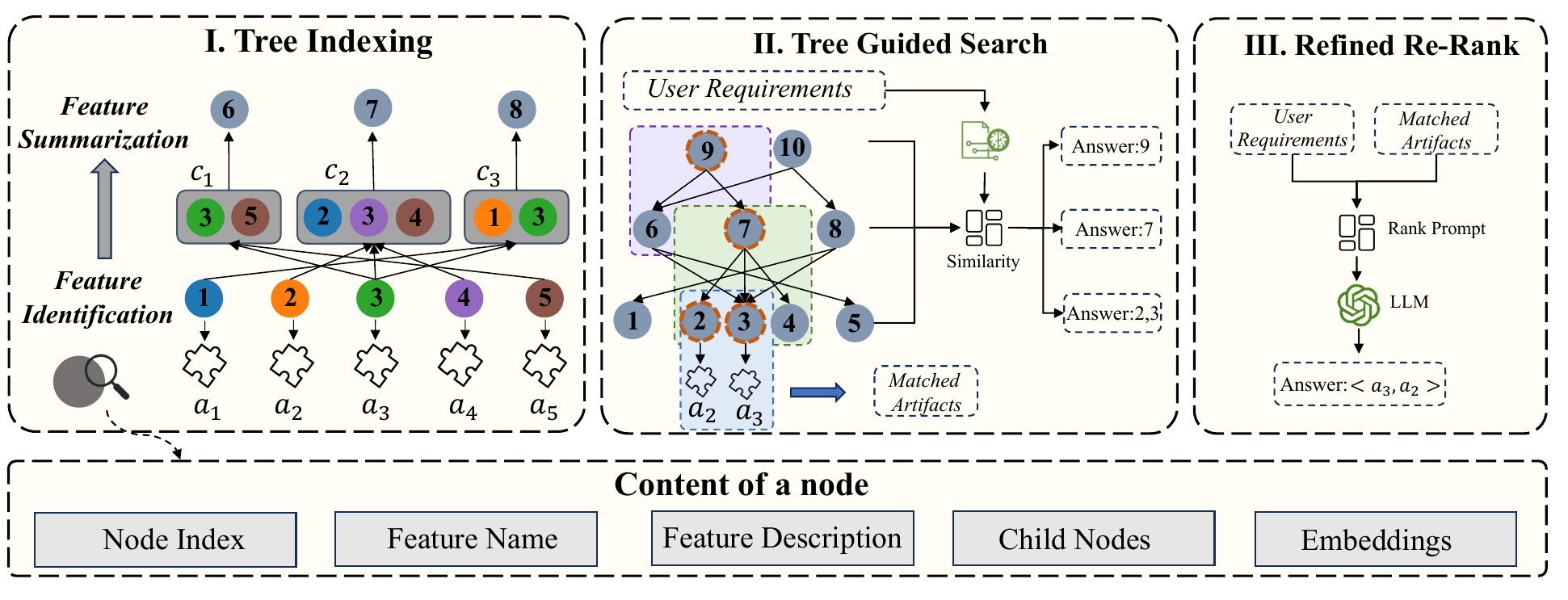}
    \caption{The Overview of \ours{}}
    \label{fig:approach}
\end{figure*}

The goal of our \ours{} is to recommend $k$ artifacts $A = \langle a_1,\dots ,a_k\rangle$ from the artifact library $L$ based on an input natural language development intent description $R$. The construction of \ours{} consists of three main steps: tree indexing, tree-guided search, and refined re-rank. The three steps work in a pipeline as shown in Figure ~\ref{fig:approach}:
\begin{itemize}
    \item \textbf{Tree Indexing.} Given an artifacts Library $L$, the step recursively identifies and abstracts high-level common features among artifacts to construct a hierarchical tree structure $T$ from bottom up. 
    \item \textbf{Tree-guided Search.} Given a development intent $R$, this step traverses the tree $T$ to judge whether each node can satisfy the requirements $R$ until reaching the leaf nodes to locate a set of candidate artifacts $A'$.
    \item \textbf{Refined Re-rank.} This step prompts LLMs to understand the user requirements $R$ and candidate artifacts $A'$, and sort these artifacts based on their suitability to the requirements $R$ to get the final top-k artifacts list $A$.
\end{itemize}

\subsection{Tree Indexing} As shown in Figure ~\ref{fig:approach}, the goal of this step is to construct a hierarchical tree to manage artifacts for domain-specific ecosystems. To achieve this goal, we view the task as a bottom-up abstraction process driven by two alternating operations, \ie feature identification and feature summarization. They are applied recursively until a single root node remains or a domain‑specific stopping criterion is met.

\textbf{Common Feature Identification.} This stage aims to identify artifacts that contains common feature from the artifact library L. \ours{} treat this procedure as a cluster task, where the main purpose is to group artifacts $a_i$ based on similarities in their functional descriptions. Specifically, \ours{} first represent the descriptions of artifacts as semantic vectors $H = [h_1,h_2,...,h_n]$ using BERT-based encoder (\ie all-mpnet-base-v2). Then \ours{} employ a clustering algorithm to group artifacts, which divides these embedding vectors $H$ into $k$ clusters $C = [C_1,C_2,...,C_k]$. The artifacts with the same cluster are considered to have a common feature. 

The selected clustering algorithm plays a key role in this step. Considering that an artifact may have multi-dimensional feature, \ours{} must employ soft clustering, where an artifact can belong to multiple clusters without requiring a fixed number of clusters. Thus, \ours{} employ Gaussian Mixture Models (GMMs) as its cluster algorithm. Besides, \ours{} employ Uniform Manifold Approximation and Projection (UMAP) to reduce the dimensionality of $H$ before clustering. Our motivation is that the high dimensionality of the origin embedding $H$ can be a challenge for GMMs, as distance metrics may behave poorly when used to measure similarity in high-dimensional spaces. The optimal number of clusters is selected via the Bayesian Information Criterion (BIC). 


\textbf{Common Feature Summarization.} This stage aims to generate the high-level common feature for each cluster $C_i$. To achieve this, \ours{} employ LLMs to understand the functional descriptions of artifacts and summarize their shared characteristics into high-level common features. Specifically, \ours{} constructs a prompt template as shown in $P_s$ based on prior knowledge. Then we pass the functional descriptions of artifacts in each cluster $C_i$ into the prompt template. GPT-4 receives the filled prompt and generates a high-level common feature $F_i$ across the artifacts in the cluster. 

\begin{Prompt}{Prompt $P_s$ for Common Feature Summarization}
Based on the following sub-features, please generate a parent common feature that can cover these sub-features. 
The sub-features are: \{\textit{the descriptions of artifacts in $C_i$}\}
Please only output the common feature in the format of 'feature name: feature description:'.
\end{Prompt}

\textbf{Recursive Construction}
\ours{} recursively applies the identification and summarization pipeline to process the newly created cluster nodes, producing high summarization until a single root remains or a pre-set stopping criteria that can be defined based on domain-specific expertise. The criteria typically include the maximum depth of the tree and the maximum number of features at the highest level. The final output is a hierarchical semantic tree $T$ whose internal nodes encode increasingly general features and whose leaves correspond to the original artifacts in $L$.

\subsection{Tree-Guided Search} 
In this section, we describe how \ours{} utilizes the hierarchical index $T$ to locate a concise set of candidate artifacts. Algorithm~\ref{alg:tree-search} presents the pseudocode of the search process.


\textbf{Search Process From Top-Down.}
The tree-guided search operates as a top-down traversal of the hierarchical semantic tree $T$. This process prunes irrelevant branches, focusing the search on regions of the tree most semantically related to the given intent $R$. Specifically, starting from the root node(s) of $T$, \ours{} first encodes the development intent $R$ into an embedding vector $E_R$. Each node $n$ in the current level is also represented by its embedding $E_n$, obtained from its textual description. The algorithm then computes the semantic similarity between $E_R$ and ${E_n}$ to select the top-$k$ most similar nodes, denoted as $S_{\text{top}}$.
If all selected nodes are leaf nodes, the search terminates and returns the corresponding artifacts as the candidate set $A’$. Otherwise, the process continues recursively by expanding the children of $S_{\text{top}}$, repeating the similarity-based selection at each level until all chosen nodes correspond to individual artifacts.

\textbf{Complexity.}  Our tree-guided search algorithm offers a significant computational advantage over traditional linear search methods. At each level of the hierarchical tree, the LLMs are queried to evaluate at most a constant number of nodes and select the $k$ most relevant ones. Since the tree's height grows logarithmically with the total number of artifacts $n$, \ie $O(logn)$, the overall time complexity of this process is $O(k \log n)$. Given that $k$ is typically a small constant(representing the number of nodes to explore at each level), this effectively simplifies to an overall time complexity of $O(\log n)$.

\begin{algorithm}[t]
\caption{Tree-Guided Artifact Search}
\label{alg:tree-search}
\begin{algorithmic}[1]
\Function{Search}{$tree, R, k$}
    \State $S_{\text{current}} \gets tree.\textit{root\_nodes}$
    \While{$S_{\text{current}} \neq \varnothing$}
        \State $E_R \gets \textsc{Encode}(R)$
        \State $E_S \gets \{\textsc{Encode}(n) \mid n \in S_{\text{current}}\}$
        \State $S_{\text{top}} \gets \textsc{TopKSim}(E_R, E_S, k)$ \Comment{select top-$k$}
        \If{\textsc{AllLeaves}$(S_{\text{top}})$}
            \State \Return $S_{\text{top}}$ \Comment{top-$k$ candidate artifacts}
        \EndIf
        \State $S_{\text{current}} \gets \bigcup_{n \in S_{\text{top}}} \textsc{Children}(n)$
    \EndWhile
    \State \Return $\varnothing$
\EndFunction
\end{algorithmic}
\end{algorithm}

\subsection{Refined Re-rank} As shown in Figure~\ref{fig:approach}, this stage aims to sort the candidate artifacts $A'$ based on their suitability to the given requirements $R$. Specifically, the tree-guided search produces a small candidate set $A' = \{a'_1,a'_2,\dots , a'_{m}\}$. In this stage, \ours{} query a LLM with the user requirements $R$ and the full list of candidates $A'$ with the prompt $P_r$, asking it to directly sort the artifacts by suitability. The top-k items in the returned order form the final recommendation $A = \langle a_1, \dots , a_k \rangle$.

\begin{Prompt}{Prompt $P_r$ for Artifact Re-rank}
Given a user requirement and a list of candidate artifacts, rank the
artifacts from best match to worst match according to how well each artifact satisfies the requirement.  

User Requirements: \textit{user requirements}

Candidate Artifacts: \textit{list of <ID, artifact description>}

Please only output the ID of the sorted artifact in a list format.
\end{Prompt}

\section{Study Design}
This section elaborates on the whole experimental setting across this work, including research questions, studied LLMs, dataset, and metrics.
\subsection{Research Questions}

\textbf{RQ1 (Effectiveness): How effective is \ours{} in recommending relevant artifacts compared with existing baselines across different ecosystems?} 
This RQ aims to evaluate the effectiveness of our proposed framework. In experiments, we select GPT-4o-turbo as the base LLM for both summarization and re-ranking. The generalization capability across different LLMs will be further examined in RQ4.

\textbf{RQ2 (Efficiency): Can \ours{} significantly reduce the time cost compared with existing baselines across different ecosystems?} 
This RQ investigates the efficiency of \ours{}. Since \ours{} prunes irrelevant branches during top-down traversal, we analyze its performance in terms of average response time and the number of LLM queries, thereby validating the scalability advantage of the tree-guided search mechanism.

\textbf{RQ3 (Robustness): How robust is \ours{} across different large language models?} 
This RQ focuses on the robustness and generalization ability of \ours{} with respect to different underlying LLMs. We evaluate whether the hierarchical structure and re-ranking strategy remain effective when substituting GPT-4o-turbo with other open-source or proprietary LLMs (\eg Qwen, DeepSeek, LLaMA), thereby validating the model-agnostic design of our framework.

\textbf{RQ4 (Ablation Study): What is the performance of \ours{} when different components are removed?} 
\ours{} consists of two core components that enhance artifact recommendation performance: (1) the Tree Indexing (TI) module and (2) the Refined Re-ranking (RR) phase. Based on the experimental setup in RQ1 and RQ2, we conduct ablation studies by incrementally adding TI and RR to a GPT-4 baseline to analyze their individual contributions across multiple ecosystems in \bench{}.

\textbf{RQ5 (Tree Quality): How does the quality of the hierarchy tree by \ours{}?} This RQ aims to examine the quality of the hierarchy tree constructed by \ours{}. In experiments, we first compute the statistics of the tree for each ecosystem, including the number of layers and nodes, and the summary length of nodes. Then we evaluate the rationalization of the tree using the silhouette score and G-value score.

\begin{table*}[]
    \centering
    \setlength{\tabcolsep}{5pt}
    \caption{%
    The effectiveness comparison of our proposed \ours{} framework with baselines across three ecosystems.}
    \begin{tabular}{lcccccccccccc}
\toprule
\multirow{2}{*}{\textbf{Solutions}} & \multicolumn{4}{c}{\textbf{JavaScript}}                                                                                                           & \multicolumn{4}{c}{\textbf{HuggingFace}}                                                                                                          & \multicolumn{4}{c}{\textbf{Linux}}                                                                                                                \\ \cmidrule{2-13} 
                                    & \multicolumn{1}{l}{\textbf{P@1}} & \multicolumn{1}{l}{\textbf{P@4}} & \multicolumn{1}{l}{\textbf{DCG@2}} & \multicolumn{1}{l}{\textbf{DCG@5}} & \multicolumn{1}{l}{\textbf{P@1}} & \multicolumn{1}{l}{\textbf{P@4}} & \multicolumn{1}{l}{\textbf{DCG@2}} & \multicolumn{1}{l}{\textbf{DCG@5}} & \multicolumn{1}{l}{\textbf{P@1}} & \multicolumn{1}{l}{\textbf{P@4}} & \multicolumn{1}{l}{\textbf{DCG@2}} & \multicolumn{1}{l}{\textbf{DCG@5}} \\ \midrule
\multicolumn{13}{c}{\textbf{Baselines}} \\ \midrule
TF-IDF&    0.24&   0.42&0.30&  0.36& 0.19&  0.36&  0.23&   0.29&    0.31&      0.62&    0.43&      0.49\\
BM25&  0.20&         0.36&    0.25&  0.30&    0.25& 0.40& 0.30& 0.34& 0.41&  0.67&  0.50&     0.58\\
LSI& 0.18&   0.32&  0.22&      0.27&  0.13&        0.21&       0.15&             0.19&      0.54&                0.82&     0.65&                  0.71\\
 JenS& 0.26& 0.45& 0.32& 0.38& 0.18& 0.30& 0.22& 0.26& 0.49& 0.81& 0.59&0.68\\
 Word2Vec& 0.02& 0.04& 0.03& 0.04& 0.05& 0.14& 0.08& 0.10& 0.11& 0.20& 0.15&0.16\\
 FastText& 0.03& 0.08& 0.04& 0.06& 0.07& 0.15& 0.09& 0.12& 0.04& 0.21& 0.09&0.14\\
GPT-4&   0.49&          0.67&  0.52&            0.55&       0.20&                 0.40&        0.20&                     0.34&    0.73&                      0.87&      0.80&    0.82\\ \midrule
\multicolumn{13}{c}{\textbf{\ours{}}} \\ \midrule
\rowcolor[HTML]{00D2CB}
GPT-4  + \ours{}& \textbf{0.58}& \textbf{0.75}& \textbf{0.64}& \textbf{0.68}& \textbf{0.28}& \textbf{0.44}& \textbf{0.33}& \textbf{0.38}& \textbf{0.78}& \textbf{0.93}& \textbf{0.85}& \textbf{0.87}\\ \midrule
\rowcolor[HTML]{00D2CB}
 Relative Improvement & \textbf{$\uparrow$ 18\%}& \textbf{$\uparrow$12\%}& \textbf{$\uparrow$23\%}& \textbf{$\uparrow$ 24\%}& \textbf{$\uparrow$40\%}& \textbf{$\uparrow$ 10\%}& \textbf{$\uparrow$ 65\%}& \textbf{$\uparrow$ 12\%}& \textbf{$\uparrow$ 7\%}& \textbf{$\uparrow$ 7\%}& \textbf{$\uparrow$ 6\%}&\textbf{$\uparrow$ 6\%}\\ \bottomrule 
\end{tabular}
    \label{tab:rq1}
\end{table*}

\subsection{Evaluation Metrics} ~\label{sec:metrics}
Following the previous studies in the recommendation systems field~\cite{gao2023know}~\cite{jiang2016rosf} and feature tree extraction field~\cite{jin2025automatic}~\cite{jin2024first}, we adopt P@k and DCG@k to assess the accuracy of the recommended artifacts, the silhouette score and Gvalue score to assess the quality of the constructed hierarchy tree. 

\textbf{Metrics for recommendation.} We employ two widely used metrics to evaluate the the accuracy of recommended artifact in RQ1, RQ3 and RQ4, \ie P@k and DCG@k.

\begin{itemize}
    \item \textbf{P@k:} measures the precision of the target artifact in the recommended top-$K$ candidate artifacts. Specifically, for a given development intent, if the top-$K$ recommended artifacts include the corresponding target artifact, the recommendation is considered successful and denoted as $\text{success}(a_m \in \text{top}K) = 1$. Otherwise, it is considered unsuccessful with $\text{success}(a_m \in \text{top}K) = 0$. The $\text{P@K}$ metric is then defined as the proportion of successful recommendations among all evaluated intents, reflecting the precision in recommending the correct artifact within the top-$K$ results:
    \begin{equation}
    \text{P@K} = \frac{1}{N} \sum_{i=1}^{N} \text{success}(a_{m_i} \in \text{top}K),
    \end{equation}
    where $N$ denotes the total number of development intents. In this paper, we adopt the P@1 and P@4. 
    \item $\textbf{DCG@K}$: is another widely used top-$K$ evaluation metric that assesses recommendation performance by considering both the relevance of recommended items and their ranking positions. Unlike $\text{P@K}$, which only checks whether the correct item appears in the top-$K$ list, $\text{DCG@K}$ emphasizes the ranking order, \ie assigning higher importance to items placed closer to the top. In other words, it receives a larger reward when the correct artifact is ranked higher. The definition of $\text{success}(a_m \in \text{top}K)$ remains consistent with the previous metric, while $\text{rank}_{a_m}$ denotes the position of the master artifact $a_m$ in the ranked list. The $\text{DCG@K}$ metric is defined as:
    \begin{equation}
    \text{DCG@K} = \sum_{i=1}^{K} \frac{\text{success}(a_{m_i} \in \text{top}K)}{\log_2(\text{rank}_{a_{m_i}} + 1)}.
    \end{equation}
    In this paper, we adopt the DCG@2 and DCG@5.

\end{itemize}

\textbf{Metrics for Tree Quality.} We employ two metrics to evaluate the quality of tree structure in RQ5, \ie silhouette score and Gvalue score.

\begin{itemize}
    \item \textbf{Silhouette Score (SS)} measures the similarity of a feature to other features under the same parent and its distinction from features under different parents~\cite{vrezankova2018different}. A higher silhouette score indicates a more coherent and well-structured feature tree. Specifically, the silhouette score is calculated as follows.

    \begin{equation}
        s(f_i) = \frac{b(f_i)- a(f_i)}{max(a(f_i),b(f_i))} 
    \end{equation}

    \begin{equation}
        S = \frac{1}{N} \sum_{i=1}^{n} s(f_i)
    \end{equation}

    where $f_i$ represents the $i$-th feature, $a(f_i)$ is the average distance between feature $f_i$ and other features under the same parent, $b(f_i)$ is the average distance between feature $f_i$ and the features in the closest parent, $s(f_i)$ is the silhouette score for feature $f_i$.
    \item \textbf{Gvalue Score (GS)} is a comprehensive metric that can be used to evaluate the feature tree. It can measure the rationality of the feature tree structure and reflect whether the parent feature covers the child features. A higher value indicates a higher-quality feature tree. The calculation method can be found in its paper~\cite{jin2024first}.
\end{itemize}

\section{Results and Analysis}

\begin{table*}[]
    \centering
    \setlength{\tabcolsep}{5pt}
    \caption{The Average Recommendation Time (s) of Each Method Across Different Ecosystems.}
    \begin{tabular}{lcccccccccccc}
\toprule
\multirow{2}{*}{\textbf{Solutions}} & \multicolumn{4}{c}{\textbf{JavaScript}}                                                                                                           & \multicolumn{4}{c}{\textbf{HuggingFace}}                                                                                                          & \multicolumn{4}{c}{\textbf{Linux}}                                                                                                                \\ \cmidrule{2-13} 
                                    & \multicolumn{1}{c}{\textbf{Mean}} & \multicolumn{1}{c}{\textbf{Std}} & \multicolumn{1}{c}{\textbf{Min}} & \multicolumn{1}{c}{\textbf{Max}} & \multicolumn{1}{c}{\textbf{Mean}} & \multicolumn{1}{c}{\textbf{Std}} & \multicolumn{1}{c}{\textbf{Min}} & \multicolumn{1}{c}{\textbf{Max}} & \multicolumn{1}{c}{\textbf{Mean}} & \multicolumn{1}{c}{\textbf{Std}} & \multicolumn{1}{c}{\textbf{Min}} & \multicolumn{1}{c}{\textbf{Max}} \\ \midrule
\multicolumn{13}{c}{\textbf{Baselines}} \\ \midrule
TF-IDF     & 0.0029& 0.0021& 0.0025& 0.0666& 0.0264& 0.0060& 0.0210& 0.0464& 0.0008& $<$0.001& 0.0007& 0.001\\
BM25       & 0.0193& 0.0053& 0.0102& 0.0776& 0.0211& 0.0074& 0.0087& 0.0537& 0.0005& 0.0001& 0.0003& 0.0006\\
LSI        & 0.0179& 0.0044& 0.0129& 0.1038& 0.0039& 0.0002& 0.0036& 0.0066& 0.0009& 0.0019& 0.0004& 0.0112\\
JenS       & 3.4352& 0.0485& 3.2750& 3.7371& 11.239& 0.0556& 11.146& 11.359& 0.0067& 0.0010& 0.0065& 0.0069\\
Word2Vec   & 0.0071& 0.0033& 0.0064& 0.0807& 0.0022& 0.0002& 0.0020& 0.0051& 0.0007& 0.0002& 0.0005& 0.0023\\
FastText & 0.0070& 0.0032& 0.0063& 0.0667& 0.0027& 0.0004& 0.0024& 0.0066& 0.0006& 0.0002& 0.0004&0.0021\\
\rowcolor[HTML]{00D2CB}
GPT-4  & 643.00& 95.300& 412.00& 912.00& 525.00& 128.70& 226.00& 742.00& 11.386& 1.9456& 9.9331& 24.578\\ \midrule
\multicolumn{13}{c}{\textbf{\ours{}}} \\ \midrule
\rowcolor[HTML]{00D2CB}
GPT-4 + \ours{}    & 4.0088& 1.0504& 1.4365& 5.0111& 12.1379& 2.4605& 3.3788& 19.2132& 4.0478& 1.0045& 2.3427& 6.4649\\ \bottomrule
\end{tabular}
    \label{tab:rq2}
\end{table*}

\textbf{RQ1: How effective is \ours{} in recommending relevant artifacts compared with existing baselines across different ecosystems?}

\textbf{Results.} Table~\ref{tab:rq1} demonstrates the precision results of \ours{} compared with all baselines (Section~\ref{sec:comparative}). ``Relative Improvement'' denotes the improvement of \ours{} against GPT-4. 

\textbf{Analyse.} \textbf{(1) \ours{} achieves substantial improvements over GPT-4 across all ecosystems.} Compared with GPT-4, \ours{} consistently yields higher accuracy on every metric.
For example, in the JavaScript ecosystem, P@1 increases by 18\% and DCG@5 increases by 24\%. These gains indicate that narrowing the candidate search space enables the LLM to focus more reliably on the most relevant artifacts. \textbf{(2) The improvements by \ours{} are especially pronounced in semantically complex ecosystems.} The JavaScript and HuggingFace ecosystems exhibit the largest boosts brought by \ours{}. For example, in the HuggingFace ecosystem, DCG@2 increases by 65\%. Such results suggest that when artifact descriptions are long, noisy, or stylistically diverse, \ours{} effectively mitigates the attention drift that often occurs in direct LLM scoring.
\textbf{(3) The Linux ecosystem shows smaller yet consistent improvements.} Although the average improvement in Linux is around 6–7\%, \ours{} still outperforms GPT-4 across all evaluated metrics. For example, in the Linux ecosystem, \ours{} delivers consistent gains on P@1, P@4, and DCG metrics. This demonstrates that even in ecosystems with concise and well-structured descriptions, integrating a hierarchical filtering mechanism provides measurable benefits.

\textbf{RQ2: Can \ours{} significantly reduce the time cost compared with existing baselines across different ecosystems?}

\textbf{Results.} Table~\ref{tab:rq2} demonstrates the experimental results of \ours{} embedded with LLMs. 

\textbf{Analyse.} \textbf{(1) \ours{} reduces the recommendation time by several orders of magnitude compared with GPT-4.} The runtime drops from hundreds of seconds to just a few seconds after applying \ours{}. For example, in the JavaScript ecosystem, the mean time decreases from 643 seconds to 4.0 seconds. This dramatic reduction shows that narrowing the scoring space before LLM inference effectively eliminates the linear-time bottleneck of exhaustive artifact scanning.
\textbf{(2) The improvement remains consistent across ecosystems with different scales.} Despite variations in corpus size and description length, \ours{} consistently accelerates GPT-4 by two to three orders of magnitude. For example, in the HuggingFace ecosystem, the mean time decreases from 525 seconds to 12.1 seconds. This demonstrates that the efficiency gains achieved by \ours{} are not tied to specific dataset characteristics but stem from its general ability to prune irrelevant candidates.
\textbf{(3) \ours{} brings LLM inference time closer to the range of traditional retrieval methods.} Although still slower than millisecond-level non-LLM baselines, the reduced runtime is now practical for interactive development workflows. For example, in the Linux ecosystem, \ours{} reduces GPT-4’s runtime from 11.4 seconds to 4.0 seconds. This suggests that hierarchical filtering not only accelerates LLM-based recommendation but also makes it feasible for real-time usage.

\begin{table*}[]
    \centering
    \setlength{\tabcolsep}{4pt}
    \caption{Experimental Results of \ours{} with Differnet LLMs}
    \begin{tabular}{lccccccccccccccc}
\toprule
\multirow{2}{*}{\textbf{LLMs}} & \multicolumn{5}{c}{\textbf{JavaScript}}& \multicolumn{5}{c}{\textbf{HuggingFace}}& \multicolumn{5}{c}{\textbf{Linux}}\\
 \cmidrule(r){2-6} \cmidrule(lr){7-11} \cmidrule(l){12-16}
 & \multicolumn{4}{c}{\textbf{Effe}}& \textbf{Effi}& \multicolumn{4}{c}{\textbf{Effe}}& \textbf{Effi}& \multicolumn{4}{c}{\textbf{Effe}}&\textbf{Effi}\\ 
 \cmidrule(r){2-6} \cmidrule(lr){7-11} \cmidrule(l){12-16}
 & \textbf{P@1}& \textbf{P@4}& \textbf{DCG@2}& \textbf{DCG@5} &\textbf{Time}& \textbf{P@1}& \textbf{P@4}& \textbf{DCG@2}& \textbf{DCG@5} &\textbf{Time}& \textbf{P@1}& \textbf{P@4}& \textbf{DCG@2}& \textbf{DCG@5} &\textbf{Time}\\ \midrule
GPT-4.1                      & 0.49& 0.67& 0.52&  0.55&643& 0.20& 0.40& 0.20&  0.34&129& 0.73& 0.87& 0.80&  0.82&11.39\\
\quad + \ours{}              & 0.58& 0.75& 0.64&  0.68&4.01& 0.28& 0.44& 0.33&  0.38&12.13& 0.78& 0.93& 0.85&  0.87&4.05\\
Improvement                  & \textcolor{red}{$\uparrow$ 18\%}& \textcolor{red}{$\uparrow$ 12\%}& \textcolor{red}{$\uparrow$ 23\%}&  \textcolor{red}{$\uparrow$ 24\%}&-& \textcolor{red}{$\uparrow$ 40\%}& \textcolor{red}{$\uparrow$ 10\%}& \textcolor{red}{$\uparrow$ 65\%}&  \textcolor{red}{$\uparrow$ 12\%}&-& \textcolor{red}{$\uparrow$ 7\%}& \textcolor{red}{$\uparrow$ 7\%}& \textcolor{red}{$\uparrow$ 6\%}&  \textcolor{red}{$\uparrow$ 6\%}&-\\ \midrule
DeepSeek-R1                  & 0.54& 0.63& 0.56&  0.56&814& 0.21& 0.32& 0.26&  0.26&449& 0.55& 0.67& 0.61&  0.63&297\\
\quad + \ours{}              & 0.58& 0.72& 0.64&  0.66&8.98& 0.22& 0.36& 0.27&  0.30&9.92& 0.61& 0.68& 0.65&  0.65&58.4\\
Improvement                  & \textcolor{red}{$\uparrow$ 7\%}& \textcolor{red}{$\uparrow$ 14\%}& \textcolor{red}{$\uparrow$ 14\%}&  \textcolor{red}{$\uparrow$ 18\%}&-& \textcolor{red}{$\uparrow$ 5\%}& \textcolor{red}{$\uparrow$ 13\%}& \textcolor{red}{$\uparrow$ 4\%}&  \textcolor{red}{$\uparrow$ 15\%}&-& \textcolor{red}{$\uparrow$ 11\%}& \textcolor{red}{$\uparrow$ 1\%}& \textcolor{red}{$\uparrow$ 7\%}&  \textcolor{red}{$\uparrow$ 3\%}&-\\ \midrule
Qwen3-8B& 0.52& 0.64& 0.56&  0.56&80.73& 0.21& 0.34& 0.21&  0.31&292.5& 0.73& 0.82& 0.79&  0.82&42.3\\
\quad + \ours{}              & 0.56& 0.75& 0.64&  0.67&1.87& 0.27& 0.41& 0.31&  0.35&4.55& 0.70& 0.78& 0.74&  0.75&2.47\\
Improvement                  & \textcolor{red}{$\uparrow$ 8\%}& \textcolor{red}{$\uparrow$ 17\%}& \textcolor{red}{$\uparrow$ 14\%}&  \textcolor{red}{$\uparrow$ 20\%}&-& \textcolor{red}{$\uparrow$ 29\%}& \textcolor{red}{$\uparrow$ 21\%}& \textcolor{red}{$\uparrow$ 48\%}&  \textcolor{red}{$\uparrow$ 13\%}&-& \textcolor{darkgreen}{$\downarrow$ 4\%}& \textcolor{darkgreen}{$\downarrow$ 5\%}& \textcolor{darkgreen}{$\downarrow$ 6\%}&  \textcolor{darkgreen}{$\downarrow$ 9\%}&-\\ \midrule
Qwen3-14B                    & 0.60& 0.70& 0.66&  0.66&44& 0.20& 0.50& 0.32&  0.36&718& 0.61& 0.69& 0.68&  0.72&80\\
 \quad + \ours{}& 0.60& 0.78& 0.67& 0.70& 2.65& 0.28& 0.45& 0.33& 0.37& 6.02& 0.56& 0.63& 0.59& 0.60&2.97\\
 Improvement& \textcolor{red}{$\uparrow$ 0\%}& \textcolor{red}{$\uparrow$ 11\%}& \textcolor{red}{$\uparrow$ 2\%}& \textcolor{red}{$\uparrow$ 6\%}& -& \textcolor{red}{$\uparrow$ 40\%}& $\downarrow$ 10\%& \textcolor{red}{$\uparrow$ 3\%}& \textcolor{red}{$\uparrow$ 3\%}& -& \textcolor{darkgreen}{$\downarrow$ 8\%}& \textcolor{darkgreen}{$\downarrow$ 9\%}& \textcolor{darkgreen}{$\downarrow$ 13\%}& \textcolor{darkgreen}{$\downarrow$ 17\%}&-\\ \midrule
 Qwen3-32B& 0.60& 0.70& 0.66& 0.66& 61& 0.20& 0.30& 0.32& 0.32& 853& 0.63& 0.68& 0.61& 0.62&35\\
\quad + \ours{}              & 0.61& 0.78& 0.68&  0.71&2.78& 0.28& 0.43& 0.34&  0.37&6.74& 0.64& 0.72& 0.67&  0.69&2.70\\
Improvement                  & \textcolor{red}{$\uparrow$ 2\%}& \textcolor{red}{$\uparrow$ 11\%}& \textcolor{red}{$\uparrow$ 3\%}&  \textcolor{red}{$\uparrow$ 8\%}&-& \textcolor{red}{$\uparrow$ 40\%}& \textcolor{red}{$\uparrow$ 43\%}& \textcolor{red}{$\uparrow$ 6\%}&  \textcolor{red}{$\uparrow$ 16\%}&-& \textcolor{red}{$\uparrow$ 2\%}& \textcolor{darkgreen}{$\downarrow$ 6\%}& \textcolor{red}{$\uparrow$ 10\%}&  \textcolor{red}{$\uparrow$ 11\%}&-\\ \bottomrule
\end{tabular}
    \label{tab:rq3}
\end{table*}

\begin{table*}[]
\centering
\setlength{\tabcolsep}{4pt}
\caption{Ablation Study of \ours{} with Different LLMs}
\begin{tabular}{lccccccccccccccc}
\toprule
\multirow{2}{*}{\textbf{LLMs}} & \multicolumn{5}{c}{\textbf{JavaScript}}& \multicolumn{5}{c}{\textbf{HuggingFace}}& \multicolumn{5}{c}{\textbf{Linux}}\\
\cmidrule(r){2-6} \cmidrule(lr){7-11} \cmidrule(l){12-16}
 & \multicolumn{4}{c}{\textbf{Effe}}&\textbf{Effi}& \multicolumn{4}{c}{\textbf{Effe}}&\textbf{Effi}& \multicolumn{4}{c}{\textbf{Effe}}&\textbf{Effi}\\ \cmidrule(r){2-6} \cmidrule(lr){7-11} \cmidrule(l){12-16}
                               & \textbf{P@1}& \textbf{P@4}& \textbf{DCG@2}& \textbf{DCG@5} &\textbf{Time}& \textbf{P@1}& \textbf{P@4}& \textbf{DCG@2}& \textbf{DCG@5} &\textbf{Time}& \textbf{P@2}& \textbf{P@5}& \textbf{DCG@2}& \textbf{DCG@5} &\textbf{Time}\\ \midrule
GPT-4                          & 0.49& 0.67& 0.52&  0.55&643& 0.20& 0.40& 0.20&  0.34&129& 0.73& 0.87& 0.80& 0.82 &11.39\\              
\quad + TI             & 0.46& 0.67& 0.54&  0.58&0.02& 0.19& 0.31& 0.22&  0.26&0.02& 0.74& 0.82& 0.78& 0.79 &0.01\\
\quad + RR                 & 0.58& 0.75& 0.64&  0.68&4.01& 0.28& 0.44& 0.33&  0.38&12.13& 0.78& 0.93& 0.85& 0.87&4.05\\ \midrule
DeepSeek-R1                    & 0.54& 0.63& 0.56&  0.56&814& 0.21& 0.32& 0.26&  0.26&449& 0.55& 0.68& 0.61&  0.63&297\\
\quad + TI             & 0.43& 0.63& 0.50&  0.55&0.02& 0.18& 0.31& 0.20&  0.25&0.02& 0.57& 0.72& 0.65&  0.67&0.01\\
\quad + RR                 & 0.58& 0.72& 0.64&  0.66&8.98& 0.22& 0.36& 0.27&  0.30&9.92& 0.61& 0.68& 0.65&  0.65&58.4\\ \midrule
Qwen3-8B& 0.52& 0.64& 0.56&  0.56&80.73& 0.21& 0.34& 0.21&  0.31&292.5& 0.73& 0.82& 0.79&  0.82&42.3\\
\quad + TI             & 0.45& 0.65& 0.52&  0.57&0.16& 0.19& 0.32& 0.23&  0.27&0.014& 0.77& 0.93& 0.85&  0.87&0.01\\
\quad + RR                 & 0.56& 0.75& 0.64&  0.67&1.87& 0.27& 0.41& 0.31&  0.35&4.55& 0.70& 0.78& 0.74&  0.75&2.47\\ \midrule
Qwen3-14B                      & 0.60& 0.70& 0.66&  0.66&44& 0.20& 0.50& 0.33&  0.37&718& 0.61& 0.69& 0.68&  0.72&80\\
\quad + TI             & 0.46& 0.66& 0.53&  0.58&0.02& 0.18& 0.32& 0.22&  0.27&0.02& 0.77& 0.93& 0.85&  0.87&0.02\\
\quad + RR                 & 0.60& 0.78& 0.67&  0.70&2.65& 0.28& 0.45& 0.33&  0.37&6.02& 0.56& 0.63& 0.59&  0.60&2.97\\ \midrule
 Qwen3-32B& 0.60& 0.70& 0.66& 0.66& 61& 0.20& 0.30& 0.32& 0.32& 853& 0.63& 0.74& 0.79& 0.80&35\\
 \quad + TI             & 0.45& 0.65& 0.52& 0.57& 0.02& 0.18& 0.31& 0.21& 0.25& 0.02& 0.77& 0.93& 0.85& 0.87&0.02\\
 \quad + RR                 & 0.61& 0.78& 0.68& 0.71& 2.78& 0.28& 0.43& 0.34& 0.37& 6.74& 0.64& 0.72& 0.67& 0.69&2.70\\ \bottomrule
\end{tabular}
\label{tab:rq4}
\end{table*}

\textbf{RQ3 (Robustness): How robust is \ours{} across different large language models?} 

\textbf{Results.} Table~\ref{tab:rq3} demonstrates the results of \ours{} embedded with various LLMs, each row of ``Improvement'' denotes the relative improvement of \ours{} against its corresponding underlying LLM. 

\textbf{Analyse.} \textbf{(1) \ours{} consistently enhances recommendation accuracy across all LLMs.} Across all these LLMs, \ours{} improves most effectiveness metrics, particularly in complex ecosystems such as JavaScript and HuggingFace. For example, for GPT-4.1 in JavaScript, P@1 increases from 0.49 to 0.58 and DCG@5 increases from 0.55 to 0.68. This shows that \ours{} reliably enhances semantic alignment regardless of the underlying model.
\textbf{(2) Weaker or noisier LLMs benefit the most from \ours{}.}
Models such as Qwen3-8B and DeepSeek-R1 gain the largest boosts, especially in more challenging ecosystems like JavaScript and HuggingFace.
For example, Qwen3-8B achieves a 17\% improvement in P@4 and a 48\% improvement in DCG@2 on HuggingFace. This suggests that \ours{} compensates for weaker reasoning ability by restricting the comparison scope to semantically relevant candidates.
\textbf{(3) \ours{} dramatically reduces recommendation latency for every LLM.}
The recommendation latency drops from hundreds of seconds to just a few seconds after applying \ours{}. For example, GPT-4.1’s average inference time in JavaScript decreases from 643 seconds to 4.01 seconds, and Qwen3-14B’s time drops from 44 seconds to 2.65 seconds. This demonstrates that \ours{} removes the linear-time bottleneck of exhaustive artifact scoring and enables near real-time inference.
\textbf{(4) The improvements are robust across architectures, model sizes, and ecosystems.} All these LLMs (\eg GPT-based and Qwen-based) benefit from \ours{} in terms of effectiveness and efficiency. This confirms that \ours{} functions as a model-agnostic enhancement layer for LLM-based recommendation.

\textbf{RQ4 (Ablation Study): What is the performance of \ours{} when different components are removed?} 

\textbf{Results.} Table~\ref{tab:rq4} demonstrates the contributions of each component (\ie TI and RR) of \ours{} on various LLMs. 

\textbf{Analyse.} \textbf{(1) Adding the Tree Indexing (TI) module brings large efficiency gains with minor changes to accuracy.} TI sharply reduces the candidate pool through hierarchical indexing, cutting inference time from hundreds of seconds to milliseconds. For example, with GPT-4 in JavaScript, latency drops from 643 seconds to 0.02 seconds while DCG@5 increases slightly from 0.55 to 0.58. This demonstrates that TI provides fast coarse filtering with largely preserved semantic quality.
\textbf{(2) Adding the Refined Re-ranking (RR) phase on top of TI substantially boosts effectiveness.} RR performs deeper LLM-guided semantic comparison on the filtered candidates. For example, GPT-4 + TI + RR achieves P@1 = 0.58 and DCG@5 = 0.68, improving over both the baseline GPT-4 and the TI-only variant.
This confirms that RR restores fine-grained semantic alignment that TI alone cannot provide.
\textbf{(3) The combined TI and RR pipeline achieves the optimal balance between precision and efficiency.} When both components are added, \ours{} reduces time costs by two to three orders of magnitude while improving all precision metrics across ecosystems. For instance, GPT-4’s latency falls from 643 seconds to 4.01 seconds while achieving consistent gains in P@k and DCG@k.
This synergy highlights that TI efficiently narrows the search space, and RR delivers high-quality semantic ranking.
\textbf{(4) The incremental addition of TI and RR provides consistent benefits across different LLMs.} General, reasoning, and open-source models all show the same upward trend when TI and RR are added sequentially. For example, Qwen3-8B improves P@4 from 0.64 to 0.75 while reducing time from 80.73 seconds to 1.87 seconds. This confirms that \ours{} is model-agnostic and scales reliably across architectures.

\begin{table}[]
    \centering
    \caption{The Statistics and quality evaluation of the tree for each ecosystem}
    \setlength{\tabcolsep}{4pt}
    \begin{tabular}{lcccccc}
\toprule
\multirow{2}{*}{\textbf{Ecosystem}} & \multicolumn{2}{c}{\textbf{Tree Content}}                                                                                                     & \multicolumn{4}{c}{\textbf{Tree Structure}}                                                                                                    \\ \cmidrule{2-7} 
                                    & \textbf{\begin{tabular}[c]{@{}c@{}}Leaf Node \\ Length\end{tabular}} & \textbf{\begin{tabular}[c]{@{}c@{}}Parent Node \\ Length\end{tabular}} & \multicolumn{1}{l}{\textbf{\#Layer}} & \multicolumn{1}{l}{\textbf{\#Node}} & \multicolumn{1}{l}{\textbf{SS}} & \multicolumn{1}{l}{\textbf{GS}} \\ \midrule
JavaScript                          &  610.34&                                                                        62.23&                                      4&                                     10550&                                 0.28&                                 0.59\\
HuggingFace                         &  830.84&                                                                        607.75&                                      4&                                     3655&                                 0.68&                                 0.63\\
Linux                               &  632.63&    56.43&                                      2&                                     172&                                 0.72&                                 0.56\\ \bottomrule
\end{tabular}
    \label{tab:rq5}
\end{table}

\textbf{RQ5: How does the quality of the tree by \ours{}?}

\textbf{Results.} Table~\ref{tab:rq4} shows the constructed tree by GPT-4 for each ecosystem from tree content and tree structure.

\textbf{Analyse.}  \textbf{(1) \ours{} effectively compresses artifact descriptions through high-level abstraction.} Across ecosystems, the parent-node summaries are substantially shorter than the original leaf-node descriptions. For these three ecosystems, the parent-to-leaf length ratios are approximately 10\%, 73\%, and 9\%, respectively. This indicates that \ours{} can generate concise hierarchical summaries, especially in ecosystems with cleaner or more structured text (JavaScript and Linux), while still maintaining reasonable abstraction, even in domains with long, heterogeneous descriptions such as HuggingFace.
\textbf{(2) \ours{} constructs compact and reasonably expressive tree structures.} JavaScript and HuggingFace form four-layer trees with thousands of nodes, whereas Linux forms a two-layer tree, reflecting the varying semantic granularity and diversity of different ecosystems. The silhouette scores show that HuggingFace and Linux trees exhibit stronger intra-cluster coherence than JavaScript. Meanwhile, the GValue scores indicate that parent-node summaries generally provide good coverage of their child nodes across ecosystems. Together, these metrics demonstrate that \ours{} produces structures that balance compactness with semantic fidelity.


\section{Disscussion}
\subsection{Case Study}

\begin{figure*}
    \centering
    \includegraphics[width=0.95\linewidth]{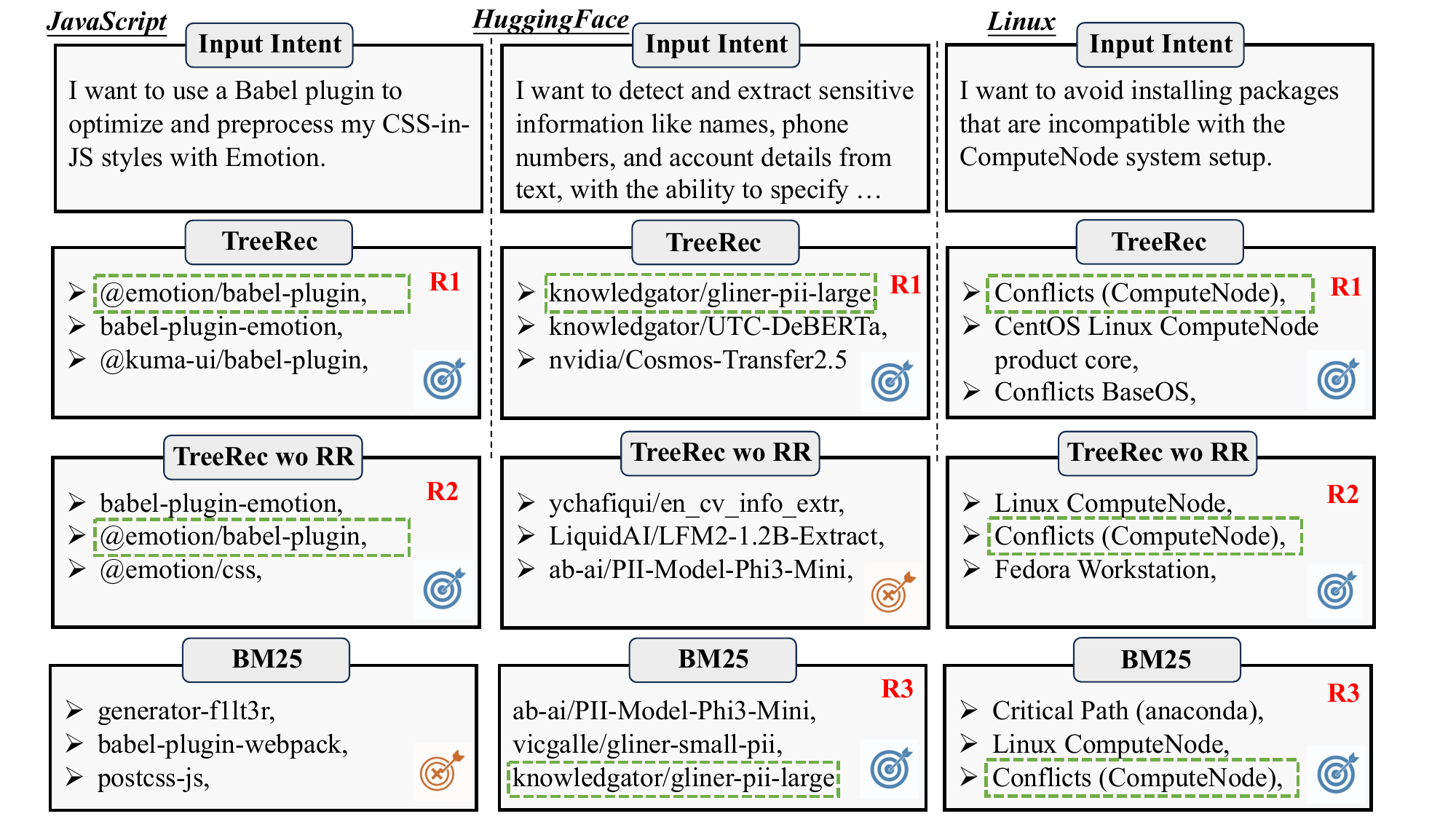}
    \caption{Case Study}
    \label{fig:case}
\end{figure*}

To complement the quantitative evaluation, we conduct a case study to qualitatively compare how different methods rank relevant artifacts.
Figure~\ref{fig:case} reports the top recommended results produced by \ours{}, \ours{} without re-ranking, and BM25 across three representative intents. The observations are consistent across all settings. First, \ours{} demonstrates the most reliable recommendation accuracy: in all three examples, it successfully retrieves all ground-truth artifacts, and notably places the correct item at the very top (Top-1) each time. This indicates that \ours{} not only identifies the correct candidates but also prioritizes them effectively, highlighting its ability to leverage hierarchical structure and latent functional clustering. Second, removing the re-ranking component leads to observable performance degradation. In one example, the method fails to include one ground-truth artifact, and even when it retrieves the correct items, they appear in second position rather than Top-1, reflecting weakened fine-grained discrimination. Third, BM25 shows clear limitations. Similar to the ablated variant, BM25 misses one ground-truth item in one case, and for the examples where it does retrieve the correct artifacts, their positions are consistently lower (Top-3 rather than Top-1). This suggests that BM25’s reliance on lexical similarity prevents it from capturing deeper structural and semantic relationships within the ecosystems. Overall, this case study confirms that \ours{} provides the most accurate and well-ranked recommendations, while both the ablated model and BM25 struggle with completeness and ranking quality.

\subsection{Threats to validity}
\textbf{Construct Validity} concerns the relationship between treatment and outcome. The threat comes from the rationality of the research questions we posed. We aim to conduct an empirical study on the performance of LLMs on reusable artifact recommendation and propose a new framework to enhance their ability. To achieve this goal, we focus on benchmark construction, empirical evaluation, performance comparison, ablation study, and tree statistics. We believe that these questions have great potential to provide insights into the ability of LLMs and illustrate the effectiveness of our proposal. 

\textbf{Internal Validity} concerns the threats to the way we perform our study. The first threat is related to the construction of our \bench{} (Section~\ref{sec:benchmark}). To construct each sample, a pair of user requirements and a corresponding artifact, we collect many artifacts from three widely used ecosystems and write user requirements by hand for each artifact. In this process, there is no threat to artifact functional description because it comes frothe m real world. However, for user requirements, we acknowledge that these creations by hand are somewhat subjective. To mitigate this threat, we conduct strict quality control before, during, and after creating user requirements (Quality Control in Section~\ref{sec:benchmark}). The second threat relates to the setting of LLMs for RQ1, RQ2, and RQ3. LLMs may show different performances under different decoding strategies or inference frameworks. To mitigate this threat, we set LLMs as the same decoding strategies and inference framework. The third threat concerns the measurement of recommendation time. Because our evaluation is conducted through third-party inference services rather than local deployments under identical hardware settings, the response time may be affected by differences in service infrastructures or optimization pipelines. This introduces potential bias in cross-LLM comparisons. Nevertheless, this setup reflects a realistic usage scenario, as many organizations rely on external API-based services rather than self-hosting LLMs. Thus, the reported latency represents the experience typical users are likely to encounter.

\textbf{External Validity} concerns the threats to generalize our findings. The first threat is the representativeness of our \bench{}. To mitigate this threat, the selected ecosystems in our benchmark are widely used and cover multiple artifact types (\ie packages, pretrained models, and groups). This ensures that the collected artifacts can represent the diversity of reusable software artifacts. Besides, the collected artifacts in our benchmark are selected based on the order of download count or popularity trend from each ecosystem. The second threat is the selection of LLMs. We only select five LLMs for our experiments. To ensure their representativeness, these models cover different parameter sizes and families (Section~\ref{sec:comparative}). 



\section{conclusion}
This work presents IntentRecBench, a large and diverse benchmark for intent-driven artifact recommendation, and provides the first comprehensive study comparing non-LLM and LLM-based solutions. Results show that while LLMs perform better than traditional methods, they still face accuracy limitations and high latency. To address these issues, we propose TreeRec, a tree-guided framework that organizes artifacts into hierarchical semantic structures for efficient and accurate retrieval. Experiments demonstrate that TreeRec consistently improves LLM performance—boosting ranking metrics by up to 65\% and reducing inference time by orders of magnitude while maintaining strong robustness across ecosystems and models. This work lays a foundation for scalable and intelligent software reuse.



\bibliographystyle{IEEEtran}
\bibliography{references}




\end{document}